\documentclass[12pt]{article}
\usepackage{amsfonts,bm,float}
\usepackage{graphicx}
\baselineskip
\offinterlineskip
\newcommand{\n}{\newline}

\newfloat{eigenvec}{htbp}{loe}

\begin{document}

\begin{center}
{\bf Spin-Hamilton Operator, Graviton-Photon Coupling
and an Eigenvalue Problem}
\end{center}

\begin{center}
{\bf Yorick Hardy$^\ast$ and Willi-Hans Steeb$^\dag$} \\[2ex]

$\ast$
Department of Mathematical Sciences, \\
University of South Africa, Pretoria, South Africa, \\
e-mail: {\tt hardyy@unisa.ac.za}\\[2ex]

$\dag$
International School for Scientific Computing, \\
University of Johannesburg, Auckland Park 2006, South Africa, \\
e-mail: {\tt steebwilli@gmail.com}\\[2ex]
\end{center}

\strut\hfill

{\bf Abstract} We solve exactly the eigenvalue problem 
for a spin Hamilton operator describing graviton-photon coupling.
Entanglement of the eigenstates are also studied.
Other spin-coupled Hamilton operators involving
spin-1 and spin-2 are also investigated and compared.
\newline



\section{Introduction}

We study the eigenvalue problem of a spin Hamilton operator given by
photon-graviton coupling. Furthermore spin Hamilton operators
for spin-$\frac12$-spin-2-spin-spin-$\frac12$ coupling and 
spin-$\frac12$-spin-1-spin-$\frac12$ couplings are also discussed. 
\newline

The photon $\gamma$ is a spin-1 particle assumed without rest mass \cite{1} 
and is described by the traceless hermitian $3 \times 3$ spin matrices
$$
p_1 = \frac1{\sqrt2} \pmatrix { 0 & 1 & 0 \cr 1 & 0 & 1 \cr 0 & 1 & 0 }, \quad
p_2 = \frac{i}{\sqrt2} \pmatrix { 0 & -1 & 0 \cr 1 & 0 & -1 \cr
0 & 1 & 0 }, \quad
p_3 = \pmatrix { 1 & 0 & 0 \cr 0 & 0 & 0 \cr 0 & 0 & -1 }
$$
with $x \leftrightarrow 1$, $y \leftrightarrow 2$, $z \leftrightarrow 3$.
The eigenvalues of these matrices are $+1$, $0$, $-1$. 
The normalized eigenvectors for $p_1$ are
$$
{\bf u}_{1,1} = \frac12 \pmatrix { 1 \cr \sqrt2 \cr 1 }, \quad
{\bf u}_{1,0} = \frac1{\sqrt2} \pmatrix { 1 \cr 0 \cr -1 }, \quad
{\bf u}_{1,-1} = \frac12 \pmatrix { 1 \cr -\sqrt2 \cr 1 }\,. 
$$
The normalized eigenvectors for $p_2$ are 
$$
{\bf u}_{2,1} = \frac12 \pmatrix { 1 \cr \sqrt2 i \cr -1 }, \quad
{\bf u}_{2,0} = \frac1{\sqrt2} \pmatrix { 1 \cr 0 \cr 1 }, \quad
{\bf u}_{2,-1} = \frac12 \pmatrix { 1 \cr -\sqrt2 i \cr -1 }\,.
$$
The normalized eigenvectors for $p_3$ is the standard basis denoted 
by ${\bf u}_{3,1}$, ${\bf u}_{3,0}$, ${\bf u}_{3,-1}$. 
We have the well-known commutation relations
$[p_1,p_2]=ip_3$, $[p_2,p_3]=ip_1$, $[p_3,p_1]=ip_2$.
The graviton $g$ is a spin-2 particle assumed without rest mass \cite{1} and 
described by the traceless hermitian $5 \times 5$ spin matrices
$$
g_1 = \pmatrix { 0 & 1 & 0 & 0 & 0 \cr 1 & 0 & \sqrt6/2 & 0 & 0 \cr
                 0 & \sqrt6/2 & 0 & \sqrt6/2 & 0 \cr
                 0 & 0 & \sqrt6/2 & 0 & 1 \cr
                 0 & 0 & 0 & 1 & 0 }, \quad
g_2 = i\pmatrix { 0 & -1 & 0 & 0 & 0 \cr 1 & 0 & -\sqrt6/2 & 0 & 0 \cr
                 0 & \sqrt6/2 & 0 & -\sqrt6/2 & 0 \cr
                 0 & 0 & \sqrt6/2 & 0 & -1 \cr
                 0 & 0 & 0 & 1 & 0 }
$$
$$
g_3 = \pmatrix { 2 & 0 & 0 & 0 & 0 \cr 0 & 1 & 0 & 0 & 0 \cr
                 0 & 0 & 0 & 0 & 0 \cr
                 0 & 0 & 0 & -1 & 0 \cr 0 & 0 & 0 & 0 & -2 }\,.
$$
The eigenvalues of these matrices are $+2$, $+1$, $0$, $-1$, $-2$.
The normalized eigenvectors for $g_1$ are
$$
{\bf v}_{1,-2} = \frac14 \pmatrix { 1 \cr -2 \cr \sqrt6 \cr -2 \cr 1 }, \,\,\,
{\bf v}_{1,2} = \frac14 \pmatrix { 1 \cr 2 \cr \sqrt6 \cr 2 \cr 1 }, 
$$
$$
{\bf v}_{1,-1} = \frac12 \pmatrix { 1 \cr -1 \cr 0 \cr 1 \cr -1 }, \,\,\,
{\bf v}_{1,1} = \frac12 \pmatrix { 1 \cr 1 \cr 0 \cr -1 \cr -1 }, \,\,\,
{\bf v}_{1,0} = \frac{\sqrt3}{\sqrt8} 
\pmatrix { 1 \cr 0 \cr -\sqrt{2/3} \cr 0 \cr 1 }\,.
$$
The normalized eigenvectors for $g_2$ are
$$
{\bf v}_{2,-2} = \frac14 \pmatrix { 1 \cr -2i \cr -\sqrt6 \cr 2i \cr 1 }, \,\,\,
{\bf v}_{2,2} = \frac14 \pmatrix { 1 \cr 2i \cr -\sqrt6 \cr -2i \cr 1 }, 
$$
$$
{\bf v}_{2,-1} = \frac12 \pmatrix { 1 \cr -i \cr 0 \cr -i \cr -1 }, \,\,\,
{\bf v}_{2,1} = \frac12 \pmatrix { 1 \cr i \cr 0 \cr i \cr -1 }, \,\,\,
{\bf v}_{2,0} = \frac{\sqrt3}{\sqrt8} 
\pmatrix { 1 \cr 0 \cr \sqrt{2/3} \cr 0 \cr 1 }\,.
$$
The normalized eigenvectors for $g_3$ are the standard basis
${\bf e}_j$ ($j=1,\dots,5$) with ${\bf v}_{3,2}={\bf e}_1$, ${\bf v}_{3,1}={\bf e}_2$,
${\bf v}_{3,0}={\bf e}_3$, ${\bf v}_{3,-1}={\bf e}_4$, 
${\bf v}_{3,-2}={\bf e}_5$. We have the well-known commutation relations
$[g_1,g_2]=ig_3$, $[g_2,g_3]=ig_1$, $[g_3,g_1]=ig_2$. 
\newline

We investigate the eigenvalue problem for the Hamilton operator
of the coupled photon-graviton spin system
$$
\hat K \equiv \frac{\hat H}{\hbar\omega} = p_1 \otimes g_1 \otimes p_1 
+ p_2 \otimes g_2 \otimes p_2
+ p_3 \otimes g_3 \otimes p_3
$$
where $\otimes$ denotes the Kronecker product \cite{2}. 
The Hamilton operator $\hat K$ acts in the Hilbert space 
${\mathbb C}^{45}$. Note that $\hat K$ is a hermitian matrix and 
thus the eigenvalues are real. Since the trace of the matrices 
$p_1$, $p_2$, $p_3$, $g_1$, $g_2$, $g_3$ vanish, we find that $\mbox{tr}(\hat K)=0$.
Consequently the sum of the 45 eigenvalues is 0.
We also study the entanglement of the the non-degenerate
eigenvectors utilizing the Schmidt decomposition
\cite{3,4,5,6,7,8}.

\section{Spectrum} 

To find an estimate for the eigenvalue we can utilize the inequality
$$
|\lambda| \le \max_{1 \le j \le n}\sum_{\ell=1}^n |a_{j\ell}|
$$
which is valid for any eigenvalue of an $n \times n$ matrix $A$.
For the Hamilton operator $\hat K=(k_{j\ell})$ we find
$$
\max_{1 \le j \le 45}\sum_{\ell=1}^{45} |k_{j\ell}| = 
4\sqrt3.
$$
A numerical study to find the eigenvalues of the Hamilton operator 
$\hat K$ using the eigenvalue packages of R \cite{9} and 
Octave \cite{10} provides the hint that 0 (7 times degenerate), 1 and $-1$ 
(each 6 times degenerate) and 2 and $-2$ (each 6 times degenerate)
are eigenvalues. Now we calculate symbolically the 
characteristic polynomial $\det(\hat K-\lambda I_{45})$. The eigenvalues
given above can now be used to reduce the characteristic polynomial
and we finally arrive at the following 45 eigenvalues
ordered from smallest to largest
\begin{eqnarray*}
\lambda_{1,2,3} &=& 
-\frac{\sqrt{\sqrt{33}+9}}{\sqrt2}, \,\,\, (3 \,\, times)\\
\lambda_{4,5,6,7,8,9} &=& -2, \,\,\, (6 \,\, times) \\
\lambda_{10} &=& -\sqrt{3} \,\,\, (1 \,\, times) \\
\lambda_{11,12,13} &=& -\frac{\sqrt{9-\sqrt{33}}}{\sqrt2} \,\,\, 
(3 \,\, times) \\
\lambda_{14,15,16,17,18,19} &=& -1 \,\,\, (6 \,\, times) \\
\lambda_{20,21,22,23,24,25,26} &=& 0 \,\,\, (7 \,\, times) \\
\lambda_{27,28,29,30,31,32} &=& 1 \,\,\, (6 \,\, times) \\
\lambda_{33,34,35} &=& \frac{\sqrt{9-\sqrt{33}}}{\sqrt2} \,\,\,
(3 \,\, times) \\ 
\lambda_{36} &=& \sqrt{3} \,\,\, (1 \,\, times) \\
\lambda_{37,38,39,40,41,42} &=& 2, \,\,\, (6 \,\, times) \\
\lambda_{43,44,45} &=& \frac{\sqrt{\sqrt{33}+9}}{\sqrt2}, \,\,\, (3 \,\, times)
\end{eqnarray*}
The eigenvalues are symmetric around 0.
Only the eigenvalues $\sqrt3$ and $-\sqrt3$ are not degenerate.
The normalized eigenvectors ${\bf w}_j$ $(j=1,\dots,45)$
are pairwise orthogonal and thus form an orthonormal basis in the 
Hilbert space ${\mathbb C}^{45}$.
\newline

Owing to the degeneracies of most of the eigenvalues the Hamilton
operator $K$ admits symmetries, i.e. 
$$
P^T\hat KP=\hat K
$$
where $P$ is a $45 \times 45$ permutation matrix. One of them is 
the permutation matrix
$$
P=\pmatrix{0&0&1\cr0&1&0\cr1&0&0}
\otimes I_5 \otimes \pmatrix{0&0&1\cr0&1&0\cr1&0&0}
$$
where $I_5$ is the $5 \times 5$ identity matrix and 
the first matrix in the Kronecker product is the
$3 \times 3$ NOT-gate, i.e. the $3 \times 3$ matrix
with all 1's in the counter-diagonal and 0 otherwise.
We have $U_{NOT}=U_{NOT}^*$ and $\hat K=P^*\hat KP$.\\

All symmetries of $\hat K$ are given by
$$
U \left(\bigoplus_{j=1}^{11} V_{j}\right) U^*
$$
and $V_1,V_2,V_3$ and $V_4$ are unitary $6\times 6$ matrices,
$V_5,V_6,V_7,V_8$ are unitary $3\times 3$ matrices,
$V_9$ and $V_{10}$ are $1\times 1$ and unitary and
$V_{11}$ is a $7\times 7$ unitary matrix, and

\begin{eqnarray*}
U&=&\sum_{j=1}^6 \mathbf{w}_{\lambda_4,j}\mathbf{e}_{j}
   +\sum_{j=1}^6 \mathbf{w}_{\lambda_{14},j}\mathbf{e}_{j+6}
   +\sum_{j=1}^6 \mathbf{w}_{\lambda_{27},j}\mathbf{e}_{j+12}
   +\sum_{j=1}^6 \mathbf{w}_{\lambda_{37},j}\mathbf{e}_{j+18}\\
&&+\sum_{j=1}^3 \mathbf{w}_{\lambda_1,j}\mathbf{e}_{j+24}
   +\sum_{j=1}^3 \mathbf{w}_{\lambda_{11},j}\mathbf{e}_{j+27}
   +\sum_{j=1}^3 \mathbf{w}_{\lambda_{33},j}\mathbf{e}_{j+30}
   +\sum_{j=1}^3 \mathbf{w}_{\lambda_{43},j}\mathbf{e}_{j+33}\\
&&+\mathbf{w}_{\lambda_{10}}\mathbf{e}_{j+36}
   +\mathbf{w}_{\lambda_{36}}\mathbf{e}_{j+37}
   +\sum_{j=1}^7 \mathbf{w}_{\lambda_{20},j}\mathbf{e}_{j+38}
\end{eqnarray*}

where $\mathbf{w}_{\lambda_k,j}$ are the elements of an orthonormal
basis for each eigenvalue $\lambda_k$ of $\hat K$, and
$\{\,\mathbf{e}_1,\,\ldots,\,\mathbf{e}_{45}\,\}$ denotes the
standard basis in $\mathbb{C}^{45}$.

\section{Entanglement}

The normalized eigenvectors of $\hat K$ are elements of the Hilbert space
${\mathbb C}^{45}$ and form an orthonormal basis in
${\mathbb C}^{45}$. Now $45=9 \cdot 5=3 \cdot 5 \cdot 3$.
The normalized eigenvectors are pairwise orthogonal and form an
orthonormal basis in ${\mathbb C}^{45}$.  
We ask the question whether the eigenvectors can be written
as the Kronecker product of vectors in ${\mathbb C}^9$ 
and vectors in ${\mathbb C}^5$. For the vector space ${\mathbb C}^9$\
we could ask again whether the vector can be written as
a Kronecker product of two vectors in ${\mathbb C}^3$.
\newline

Consider the eigenvectors belonging to the non-degenerate
eigenvalues $\pm\sqrt3$ (page \pageref{eigenvec}).%
\begin{eigenvec}
{\footnotesize%
$$
\mathbf{w}_{\sqrt3}=\frac{1}{\sqrt{24(2-\sqrt3)}}\pmatrix{
0 \cr 0 \cr 0 \cr 1 \cr 0 \cr (2-\sqrt3) i \cr 0 \cr 0 \cr 0 \cr (\sqrt3 - 2) i \cr 0 \cr - 1 \cr
0 \cr 0 \cr 0 \cr 0 \cr (\sqrt3-1)(i+1) \cr 0 \cr 0 \cr 0 \cr 0 \cr 0 \cr 0 \cr 0 \cr 0 \cr
0 \cr 0 \cr 0 \cr (1-\sqrt3)(i+1) \cr 0 \cr 0 \cr 0 \cr 0 \cr (2-\sqrt3) i \cr 0 \cr
1 \cr 0 \cr 0 \cr 0 \cr -1 \cr 0 \cr (\sqrt3 - 2) i \cr 0 \cr 0 \cr 0},\quad
\mathbf{w}_{-\sqrt3}=\frac{1}{\sqrt{24(2+\sqrt3)}}\pmatrix{
0 \cr 0 \cr 0 \cr 1 \cr 0 \cr (\sqrt3 + 2) i \cr 0 \cr 0 \cr 0 \cr -(\sqrt3 + 2) i \cr 0 \cr - 1 \cr
0 \cr 0 \cr 0 \cr 0 \cr -(\sqrt3 + 1) (i+1) \cr 0 \cr 0 \cr 0 \cr 0 \cr 0 \cr 0 \cr 0 \cr 0 \cr
0 \cr 0 \cr 0 \cr (\sqrt3 + 1)(i+1) \cr 0 \cr 0 \cr 0 \cr 0 \cr (\sqrt3 + 2) i \cr 0 \cr 1 \cr
0 \cr 0 \cr 0 \cr -1 \cr 0 \cr -(\sqrt3 + 2) i \cr 0 \cr 0 \cr 0}.
$$}
\strut\\
\centerline{Eigenvectors for $\sqrt3$ and $-\sqrt3$.}
\label{eigenvec}
\end{eigenvec}
We find the Schmidt decompositions
\begin{eqnarray*}
\lefteqn{\sqrt{24(2-\pm\sqrt3)}\mathbf{w}_{\pm\sqrt3}^T}\qquad&&\\
&=& P_{GB} \pmatrix{1,\,0,\,0,\,0,\,0,\,0,\,0,\,0,\,1}\otimes\pmatrix{0,\,1,\,0,\,(\pm\sqrt{3}-2)i,\,0}\\
&&+ P_{GB} \pmatrix{0,\,0,\,1,\,0,\,0,\,0,\,1,\,0,\,0}\otimes\pmatrix{0,\,(2-\pm\sqrt{3})i,\,0,\,-1,\,0}\\
&&+ (1-\pm\sqrt{3})(i+1)P_{GB} \pmatrix{0,\,0,\,0,\,0,\,1,\,0,\,0,\,0,\,0}
\otimes\pmatrix{-1,\,0,\,0,\,0,\,1}
\end{eqnarray*}
(where $P_{GB}$ is the permutation matrix which rearranges the
photon-photon-graviton state as a photon-graviton-photon state)
which expresses the (photon pair) -- (graviton) entanglement,
\begin{eqnarray*}
\lefteqn{\sqrt{24(2-\pm\sqrt3)}\mathbf{w}_{\pm\sqrt3}^T}\qquad&&\\
&=& \pmatrix{0,\,(\pm\sqrt{3}-1)(i+1),\,0}
    \otimes\pmatrix{0,\,1,\,0,\,0,\,0,\,0,\,0,\,0,\,0,\,0,\,0,\,0,\,0,\,-1,\,0}\\
&&+ \pmatrix{1,\,0,\,(2-\pm\sqrt{3})i,\,0}
    \otimes\pmatrix{0,\,0,\,0,\,1,\,0,\,0,\,0,\,0,\,0,\,0,\,0,\,-1,\,0,\,0,\,0}\\
&&+ \pmatrix{(2-\pm\sqrt3)i,\,0,\,1}
      \otimes\pmatrix{0,\,0,\,0,\,0,\,0,\,1,\,0,\,0,\,0,\,-1,\,0,\,0,\,0,\,0,\,0}
\end{eqnarray*}
which expresses the (first photon) -- (graviton - second photon) entanglement, and
\begin{eqnarray*}
\lefteqn{\sqrt{24(2-\pm\sqrt3)}\mathbf{w}_{\pm\sqrt3}^T}\qquad&&\\
&=& \pmatrix{0,\,1,\,0,\,0,\,0,\,0,\,0,\,0,\,0,\,0,\,0,\,0,\,0,\,-1,\,0}
    \otimes\pmatrix{1,\,0,\,(2-\pm\sqrt3)i}\\
&&+ \pmatrix{0,\,0,\,0,\,1,\,0,\,0,\,0,\,0,\,0,\,0,\,0,\,-1,\,0,\,0,\,0}
    \otimes\pmatrix{(\pm\sqrt3-2)i,\,0,\,-1}\\
&&+ \pmatrix{0,\,0,\,0,\,0,\,0,\,1,\,0,\,0,\,0,\,-1,\,0,\,0,\,0,\,0,\,0}
    \otimes\pmatrix{0,\,(\pm\sqrt3-1)(1+i),\,0}
\end{eqnarray*}
which expresses the (first photon - graviton) -- (second photon) entanglement.

\section{Other Spin Couplings}

For completeness we also provide the eigenvalues for
spin-$\frac12$-spin-2-spin-$\frac12$ and 
spin-$\frac12$-spin-1-spin-$\frac12$ couplings.
The spin matrices $s_1$, $s_2$, $s_3$ for spin-$\frac12$ are given 
by
$$
s_1 = \frac12 \pmatrix { 0 & 1 \cr 1 & 0 }, \quad
s_2 = \frac12 \pmatrix { 0 & -i \cr i & 0 }, \quad
s_3 = \frac12 \pmatrix { 1 & 0 \cr 0 & -1 }
$$
with the commutation relations $[s_1,s_2]=is_3$,
$[s_2,s_3]=is_1$, $[s_3,s_1]=is_2$.
We consider the spin-Hamilton operators
$$
\hat K_{ge} = \frac{\hat H_{ge}}{\hbar\omega} = 
s_1 \otimes g_1 \otimes s_1 + s_2 \otimes g_2 \otimes s_2 + s_3 \otimes g_3 \otimes s_3
$$
and
$$
\hat K_{pe} = \frac{\hat H_{pe}}{\hbar\omega} = 
s_1 \otimes p_1 \otimes s_1 + s_2 \otimes p_2 \otimes s_2 
+ s_3 \otimes p_3 \otimes s_3\,.
$$
The Hamilton operators $\hat K_{ge}$ for the 
spin-$\frac12$-spin-2-spin-$\frac12$ coupling 
is a hermitian $20 \times 20$ matrix with trace equal to 0. 
The eigenvalues are
$$
-\frac{\sqrt3}{4}, \quad \frac{\sqrt3}{4}, \quad -\frac{\sqrt3}{2}, \quad
\frac{\sqrt3}{2}, \quad 0
$$
each with multiplicity 4. The Hamilton operators $\hat K_{pe}$ for 
the spin-$\frac12$-spin-1-spin-$\frac12$ coupling 
is a hermitian $12 \times 12$ matrix with trace equal to 0. 
The eigenvalues are
$$
-\frac{\sqrt3}{4}, \quad \frac{\sqrt3}{4}, \quad 0
$$
each with multiplicity 4. Thus we see that the eigenvalues of the 
Hamilton operator $\hat K_{pe}$ are also eigenvalues of $\hat K_{ge}$ 
with the same multiplicity and $\hat K_{ge}$ has two additional 
eigenvalues $-\sqrt3/2$ and $\sqrt3/2$ which are twice the eigenvalues 
$-\sqrt3/4$ and $\sqrt3/4$, respectively. 
The coupled photon-graviton Hamilton operator admits
the non-degenerate eigenvalues $\sqrt3$ and $-\sqrt3$
which are twice the eigenvalues $\sqrt3/2$, $-\sqrt3/2$,
respectively.
  
\section{Conclusion}

We considered a spin Hamilton for graviton-photon coupling. The 
eigenvectors of the associated Hamilton operator, for non-degenerate
eigenvalues, yield pairwise entangled systems.
Entanglement in the eigenspaces for non-degenerate eigenvalues
has not been investigated. Also the dynamics of the entanglement
under this Hamilton operator can still be investigated.
The eigenvalues for two other related spin-coupled 
Hamilton operators have been provided and compared.

\section*{Acknowledgments}

The authors are supported by the National Research Foundation (NRF),
South Africa. This work is based upon research supported by the National
Research Foundation. Any opinion, findings and conclusions or recommendations
expressed in this material are those of the author(s) and therefore the
NRF do not accept any liability in regard thereto.

\strut\hfill

\end{document}